\documentclass [aps,prl,floats,amsmath,twocolumn]{revtex4-1}
\usepackage{mathrsfs}
\usepackage{natbib}

\usepackage{graphicx}						
\usepackage[english]{babel}
\usepackage[T1]{fontenc}
\usepackage[latin1]{inputenc}
\usepackage[babel=true]{csquotes}
\usepackage{color}
\usepackage{bm}
\usepackage{import}
\usepackage{dcolumn}

\begin{document}
\title {Flow-Induced Draping}
\author{Lionel Schouveiler}
\email{lionel@irphe.univ-mrs.fr}
\author{Christophe Eloy}
\affiliation{Aix-Marseille University, IRPHE UMR 7342, CNRS, 13013 Marseille, France}
\date{\today}
\begin{abstract}
Crumpled paper or drapery patterns are everyday examples of how elastic sheets can respond to external forcing. In this Letter, we study experimentally a novel sort of forcing. We consider a circular flexible plate clamped at its center and subject to a uniform flow normal to its initial surface. As the flow velocity is gradually increased, the plate exhibits a rich variety of bending deformations: from a cylindrical taco-like shape, to isometric developable cones with azimuthal periodicity two or three, to eventually a rolled-up period-three cone. We show that this sequence of flow-induced deformations can be qualitatively predicted by a linear analysis based on the balance between elastic energy and pressure force work.
\end{abstract}
\maketitle


Bending and stretching are the two deformation modes of thin elastic plates. For plates of thickness $h$ and typical length $R$, the ratio between bending and stretching energies scales as $\left(h/R\right)^2$. It results that thin plates, for which $h\ll R$, will favor bending to stretching if permitted by the boundary conditions. For instance, a suspended piece of fabric generally exhibits draperies that have a conical shape. These particular deformations are almost everywhere isometric, i.e. the surface is not stretched nor compressed compared to its initial state. In this example, fold number and size are determined by the competition between gravity and elasticity \cite{Cerda04} and stretching is focussed in a small area in the vicinity of the cone tip. Such conical singularities are referred to as d-cone, for ``developable cone''. They are one of the two kinds of elementary singularities in thin plate deformations, the other being ridges, in which stretching is focussed along lines. These two elementary singularities can be visualized by unfolding a sheet of crumpled paper: they appear respectively as crescent- or line-like permanent marks and result from plastic deformations of the paper \cite{Audoly10}.

Developable cones have been the subject of specific theoretical analyses \cite{BenAmar97}, as well as experiments consisting generally in pushing a flat elastic plate into a ring \cite{Cerda98,Chaieb98}. In addition to crumpled paper and gravity-induced draperies, d-cones have also been reported in delamination processes \cite{Chopin08}, tissue growth \cite{Dervaux08}, and in suspended thin layers of  viscous fluid \cite{Boudaoud01}. In this Letter,  we propose a novel experiment in which d-cones are induced by flow-induced loads on a thin elastic plate. In this experimental setup, a rich variety of deformations can be observed: cylindrical taco-like shapes, d-cones with two- or three-fold azimuthal symmetries, and a three-fold cone that is itself folded, thereby breaking the azimuthal periodicity.

The deformation of thin flexible structures due to flow-induced loads has been investigated in the past as a possible strategy to reduce the drag, comparatively to rigid structures \cite{Alben02,Alben04,Gosselin10}. This phenomenon has been particularly observed for plants that adopt more streamlined shapes to withstand high winds or water currents \cite{Vogel94,Langre12}. For instance, tulip trees have broad leaves that can reconfigure into cones, thus allowing them to reduce their drag and resist breakage \cite{Vogel89}.
This particular reconfiguration has motivated analytical \cite{Alben10} as well as experimental \cite{Schouveiler06} studies that have considered circular flexible plates cut along one radius, clamped at their center, and placed in a uniform flow. These plates have been shown to roll up into circular cones becoming sharper as the flow velocity is increased. In addition, the drag experienced by these plates was shown to be smaller than for rigid bodies.

\begin{figure}[b]
\includegraphics[scale=0.405]{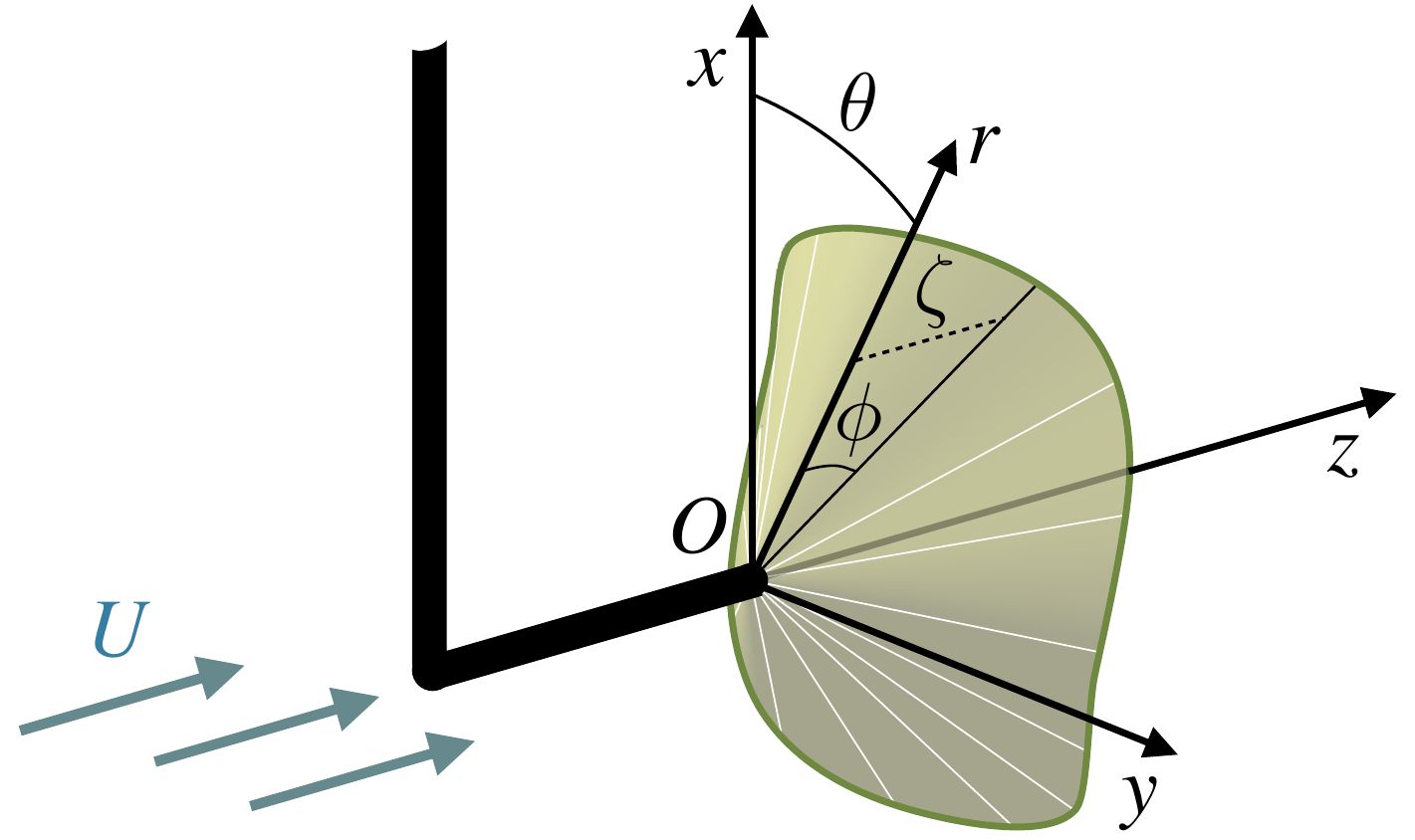}
\caption{Experimental setup. A circular silicon plate is clamped onto an elbow and placed in a uniform water flow. The function $\phi(\theta)$ describe the d-cone deformation of the plate as flow velocity $U$ is varied.}
\label{fig:sketch}
\end{figure}
\begin{figure*}[t]
\includegraphics[scale=.99]{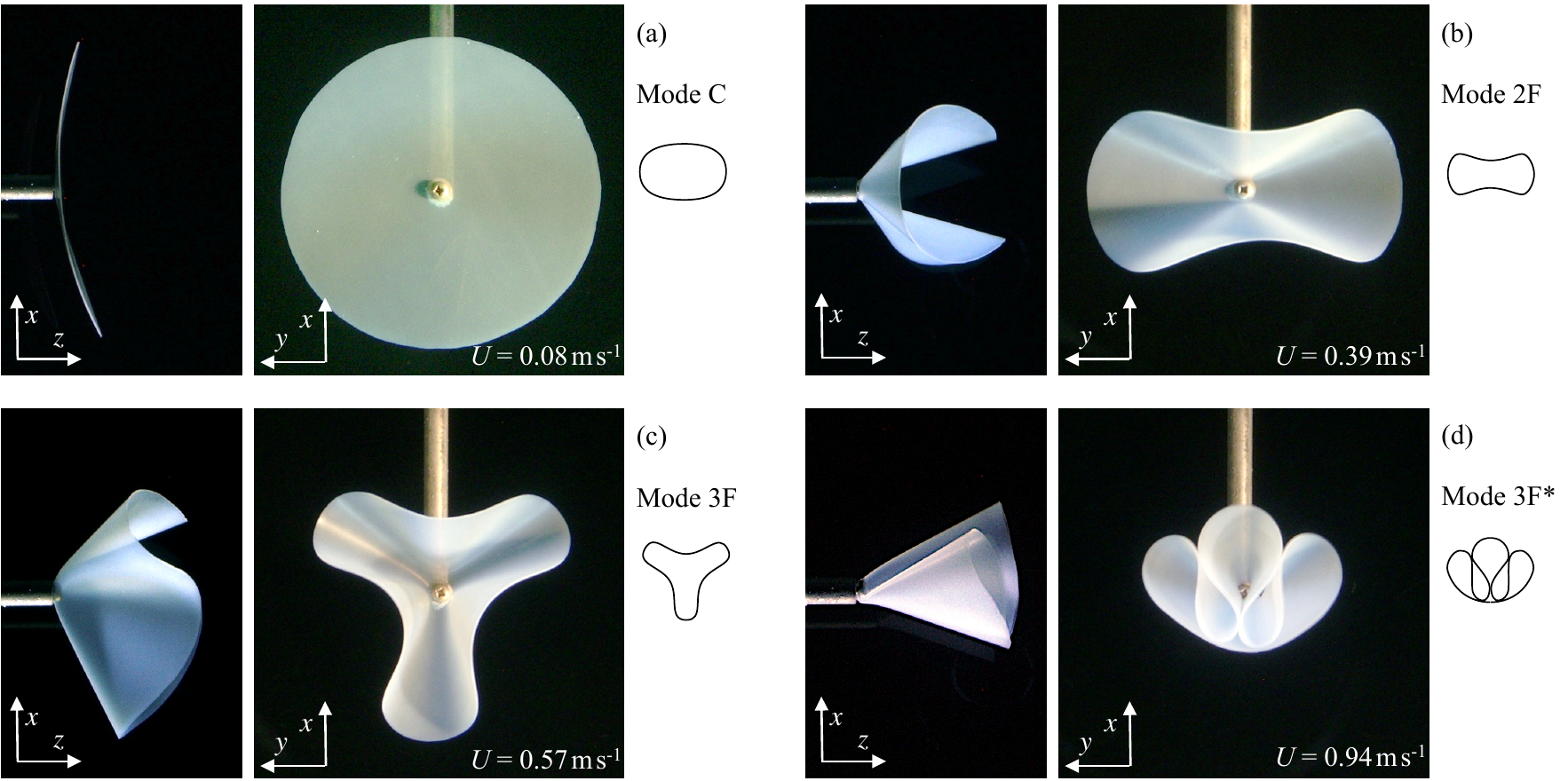}
\caption{Draping modes of a circular plate  of radius $R=0.05\,$m. When the flow velocity is gradually increased, different draping  patterns become visible starting from ``mode C'', a cylindrical mode (a), to ``mode 2F'', a two-fold conical mode (b), ``mode 3F'', a three-fold mode (c), and eventually ``mode 3F*'', a bent three-fold mode (d). }
\label{fig:pictures}
\end{figure*}

%
%
In the present study, experiments are performed with  circular silicon plates of thickness $h=5\times10^{-4}\,$m, bending rigidity $B=180\times10^{-6}\,$N$\,$m, density $\rho_p=1200\,$kg$\,$m$^{-3}$, and variable radius $R$. The plate is attached at its center to an upstream elbow and held in a horizontal free-surface water channel of $0.38\,$m-width and $0.45\,$m-high test section (Fig.~\ref{fig:sketch}).
The plate is placed in a steady uniform flow of velocity $U$ and density $\rho=1000\,$kg$\,$m$^{-3}$, perpendicular to the plane that initially contains the plate.

In these experiments, we have varied two control parameters: the plate radius $R$ in the interval $0.03<R<0.1\,$m, and the flow velocity $U$ up to $1\,$m$\,$s$^{-1}$.
A typical experiment consists, for a given plate of radius $R$, of gradually increasing the flow velocity starting from $U=0$. As the plate deforms due to flow-induced loads, it is imaged through the transparent channel walls, both along the $y$- and $z$-axes (Fig.~\ref{fig:sketch}).
Silicon has been chosen as the plate material because it allows us to have a  flat plate in the absence of flow thanks to the plate rigidity  and to the silicon density relative to water that is close to 1, such that buoyancy is negligible.

%
%
For each plate radius studied, the sequence of deformation modes is identical. At small flow velocities, the plate bends around one of its diameters forming a cylindrical surface with generatrices perpendicular to the incoming flow (Fig.~\ref{fig:pictures}a). This mode has been called ``mode C'', for cylindrical. This sort of bending deformation has  been extensively studied in the literature, both experimentally and theoretically. Studies have considered three-dimensional geometries such as rectangular plates in  uniform flow \cite{Gosselin10} or two-dimensional geometries in which a flexible filament was placed into a flowing soap film \cite{Alben02, Alben04}.

For larger flow velocities, different draping modes appear. These draping patterns are shown in Fig.~\ref{fig:pictures}b--d. They have approximately a conical shape that can be analytically described by the angle $\phi(\theta)$ (Fig.~\ref{fig:sketch}).
As the flow velocity is increased, a sequence of three different conical modes can be observed. Beyond a first threshold, the plate reconfigures into a cone with $n=2$ folds (Fig.~\ref{fig:pictures}b). After a second threshold, the plate exhibits $n=3$ folds (Fig.~\ref{fig:pictures}c). Both of these deformation are azimuthally periodic such that the function $\phi(\theta)$ has a period $2\pi/n$. This periodicity is lost beyond the last threshold. The last deformation mode still presents three-folds but these folds are themselves ``rolled-up'' or pushed against each other, breaking the periodicity (Fig.~\ref{fig:pictures}d). This last mode persists up to the highest flow velocity explored during present study (i.~e. $U=1\,$m$\,$s$^{-1}$).
These conical modes, which appear in sequencial order after the ``mode C'', are called mode 2F, mode 3F, and mode 3F* respectively.
It should be pointed out that, although the elbow maintaining the plate in the flow induces a breaking of the azimuthal symmetry in the problem, the azimuthal phase of the different modes presented in Fig.~\ref{fig:pictures}a--d is arbitrary and has been observed to change when repeating the experiments.

\begin{figure}[t]
\includegraphics[scale=0.5]{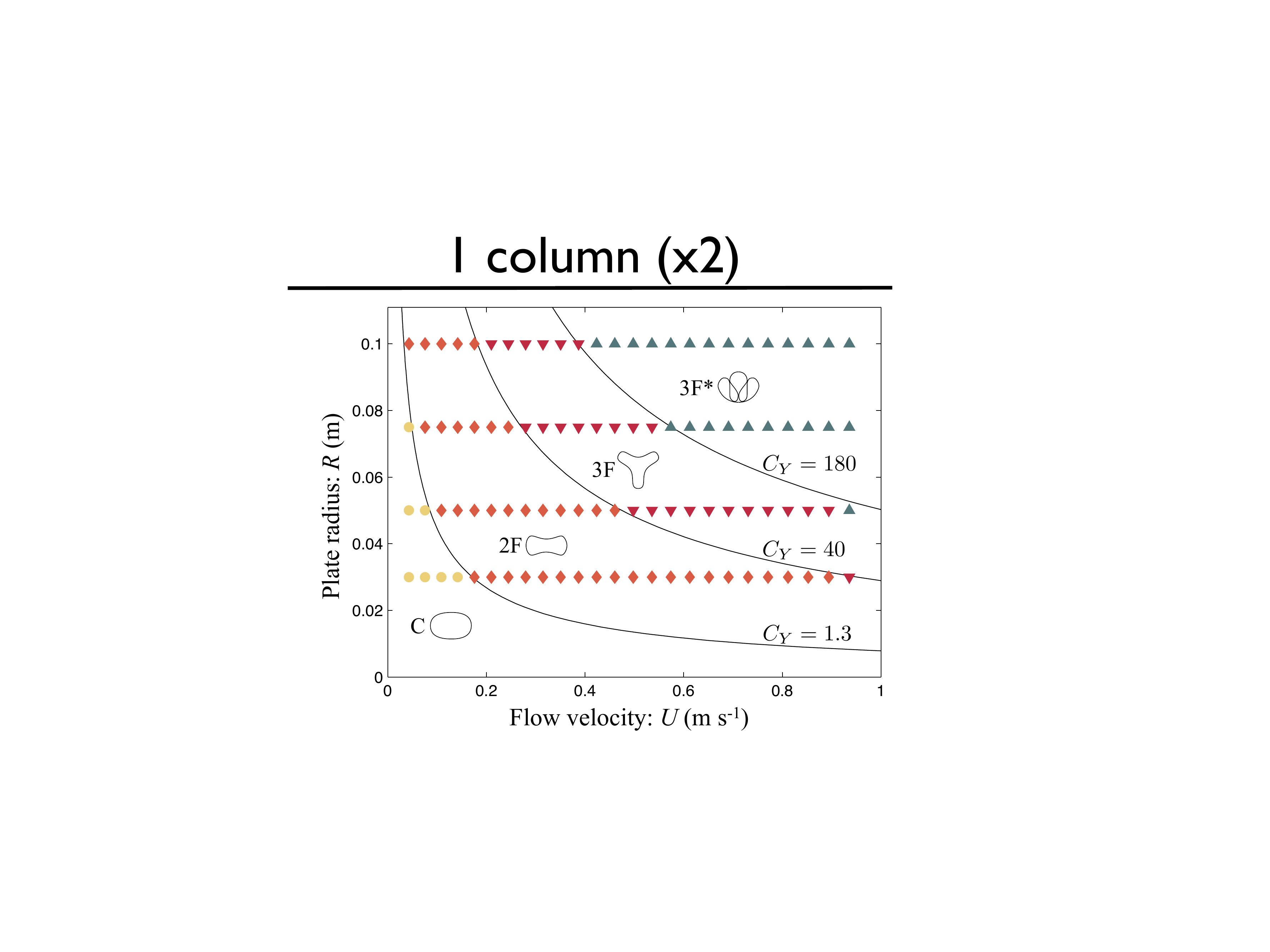}
\caption{Phase diagram in the $R$--$U$ plane. Each symbol correspond to  a different deformation mode as pictured. The solid lines correspond to iso-values of the Cauchy number as defined in Eq. (\ref{e:N_number}).}
\label{fig:phase}
\end{figure}

To assess the influence of the plate radius on the velocity thresholds, different experiments have been conducted with different radii: $R=0.03$, $0.05$, $0.075$ and $0.1\,$m. Qualitatively, the larger is the plate radius, the smaller are the thresholds.
The domains of observation of the four deformation modes in the plane $U$--$R$ are reported in a phase diagram (Fig.~\ref{fig:phase}). This diagram corresponds to experiments performed when the velocity $U$ is gradually increased and only reports modes that  develop spontaneously.
Note that, when the flow velocity is decreased, the reverse sequence of  modes can be observed (i.e. modes 3F*, 3F, 2F, and C), but the velocity thresholds are systematically lower than for increasing velocity, revealing hysteresis loops. These hysteresis loops can be as large as 30\% of the velocity thresholds and are likely due to energetic barriers between modes.

The draping patterns observed in our experiments can be considered as d-cones that result from bending deformations almost everywhere except in the vicinity of the cone tip where stretching is localized. To evidence this stretching zone, we conducted  complementary experiments with plates made of  plastic (PVC). In this case, when the plate has been deformed into a mode 2F, stretch focussing is revealed by two crescent-shape permanent marks near the tip. These marks are due to plastic deformations, when the elastic limit of the material is exceeded. The same crescent-shape singularities have been reported in  d-cones obtained by pushing a flexible circular plate into a ring of smaller diameter \cite{Chaieb98, Cerda98}. Note that, when these plastic (irreversible) deformations occur, the plate remains locked in the mode 2F, even for large flow velocity.

%
%
To gain better insight into the sequence of draping modes observed during experiments, we now consider the plate energy associated with the flow-induced conical deformations. For this purpose, we consider a thin circular sheet of bending rigidity $B$, radius $R$, area $S$ and thickness $h$ that initially lies in the plane $z=0$. When subject to a flow of density $\rho$ and velocity $U$ in the $z$-direction, this sheet deforms into a conical surface, such  that the streamwise plate deflection can be written
$\zeta(r,\theta)=r f(\theta)$,
where $(r,\theta,z)$ are the cylindrical coordinates with the origin $O$ at the cone tip (Fig.~\ref{fig:sketch}).
The function $f(\theta)$ describes the conical modes and satisfies $f(\theta)=\tan \phi(\theta)$.

In this problem, a Reynolds number, $\mbox{Re}=UR/\nu$, can be defined, which compares inertia to viscous forces, with $\nu$ the kinematic viscosity of the fluid. This Reynolds number being of the order of $10^5$, viscous effects can be neglected here. Since buoyancy can also be neglected because the plate material has almost the same density as the fluid one, the total energy of the plate $\mathcal{E}$ is the difference between its elastic energy and the potential energy corresponding to the work done by the flow pressure forces.
In the asymptotic limit of thin plates ($h/R\ll1$), we will assume that the plate undergoes pure bending almost everywhere such that elastic energy  is dominated by  bending energy:
\begin{equation}
\mathcal{E}_b=\frac{B}{2}\int_{S}\left(\Delta\zeta\right)^2 \mathrm{d}S=\frac{B}{2}\ln\frac{R}{R_c}\int_0^{2\pi}\left(f+f''\right)^2\mathrm{d}\theta,
\label{e:bending_energy}
\end{equation}
where $\Delta\zeta=\left(f+f''\right)/r$ is the local curvature of the plate and primes denote derivatives with respect to $\theta$. The cut-off radius $R_c$ in Eq.~(\ref{e:bending_energy}) comes from the logarithmically divergent integral as $r$ goes to 0. Physically this means that pure bending deformations cannot persist up to the cone tip ($r=0$). Around the tip, in a region of typical size $R_c\ll R$, stretching cannot be neglected anymore. Yet, the precise value of $R_c$ is not crucial since the bending energy $\mathcal{E}_b$ varies only logarithmically with $R_c$.

The work done by the fluid pressure when the plate deforms into a cone can be evaluated as
\begin{equation}
\mathcal{E}_p=\int_{S}\int_{0}^{\phi}pr\,\mathrm{d}\varphi\,\mathrm{d}S,
\label{e:pressure_energy}
\end{equation}
where $p$ is the local pressure jump across the plate.
With the assumptions made above, a dimensionless number can be defined, which compares the typical work done by pressure forces ($\rho U^2R^3$) to the typical bending energy ($B\ln(R/R_c)$). This elasto-hydrodynamic number is a Cauchy number and can be written as
\begin{equation}
C_Y=\frac{\rho U^2 R^3}{B\ln(R/R_c)}.
\label{e:N_number}
\end{equation}
This Cauchy number is the only dimensionless number of the problem. Iso-values of $C_Y$ are reported on the phase diagram in Fig.~\ref{fig:phase} and it  shows that the experimentally measured thresholds all correspond to a fixed value of $C_Y$. The transition between modes C and 2F occurs at $C_Y\approx1.3$, while transition between modes 2F and 3F appears at $C_Y\approx40$, and the one between 3F and 3F* at $C_Y\approx180$.

To go further, we have calculated the energy associated with linear draping modes with azimuthal wavenumbers $n= 2$, $3$, and $4$ (noted modes 2F, 3F, and 4F by analogy with experimentally observed modes). In the limit of small deflections, the developability condition reads $\int_0^{2\pi}\left(f^2-f'^2\right)\mathrm{d}\theta=0$ \cite{Audoly10}.
The simplest conical surfaces with azimuthal wavenumber $n$ that satisfy this condition are given by \cite{BenAmar97, Boudaoud01} $f(\theta)=f_0(1+\sqrt{2/(n^2-1)}\sin n\theta)$ where $f_0$ is a small amplitude. To evaluate the pressure jumps associated to these deformations, we assumed that the pressure is constant along a generatrix of the conical shape and is equal to the pressure on a circular cone of same angle as determined in \cite{Schouveiler06} using momentum conservation arguments. It yields a pressure jump, $p=\rho U^2 (1-\sin\phi)$.

To calculate the total energy of cylindrical deformations (mode C), we assumed a constant curvature everywhere on the plate, and a pressure jump, $p=\rho U_n^2$, with $U_n$ the component of the flow velocity normal to the plate surface. Note that, in order to compare cylindrical mode (for which there is no stretching) and conical modes (for which stretching is localized in a zone of size $R_c$), we used $\ln(R/R_c) = 4$. This value has been considered as a fitting parameter and corresponds  to $R_c\approx1\,$mm, which is approximately the diameter of the central hole in the plate.

\begin{figure}[t]
\includegraphics[scale=0.5]{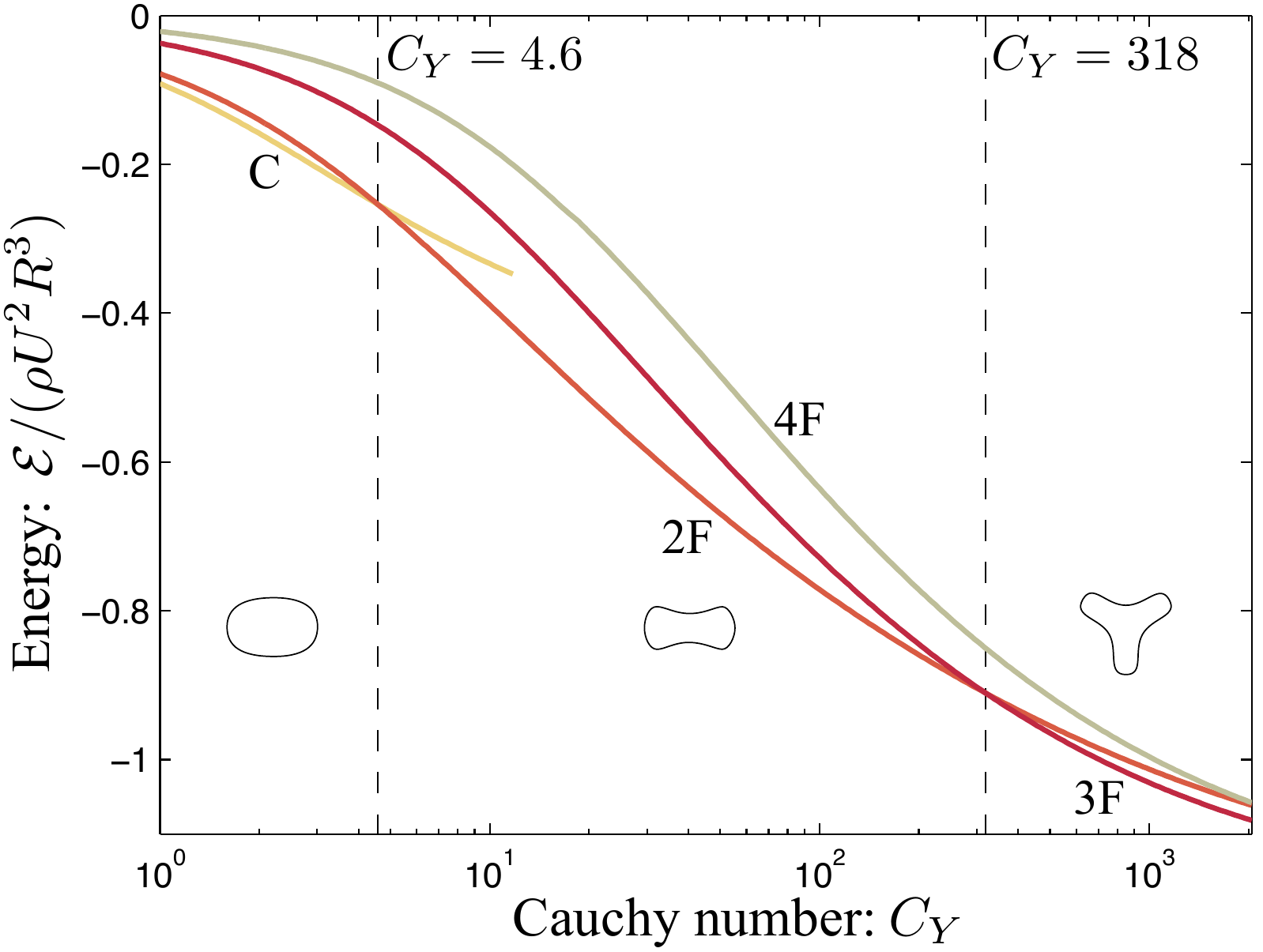}
\caption{Minimal energy as a function of the Cauchy number for the modes C, 2F, 3F, and 4F. The calculation for mode C is stopped when the radius of curvature reaches $2R/\pi$.}
\label{fig:energy}
\end{figure}
For a given Cauchy number, we can thus calculate the deformation amplitude that minimizes the total energy for each mode: C, 2F, 3F, and 4F. We can then compare these minimal energies as the Cauchy number is increased, or equivalently as the flow velocity is increased (Fig.~\ref{fig:energy}).
Although this analysis is limited by the rather crude modeling of the pressure and by the assumption of linear deflections, the modes that minimize the energy follow the same sequence as in the experiments: mode C for $C_Y<4.6$, mode 2F for $4.6<C_Y<318$, and mode 3F for $C_Y>318$.
Note that the mode with azimuthal wavenumber $n=4$ never appears to have the lowest energy. And indeed, in the experiments, even when we tried to force a 4-fold mode, the system always returned to its spontaneous two- or three-fold mode.

The above approach does not allow one to model the complex deformations observed for the largest values of $C_Y$ (mode 3F*). However, for this mode, a qualitative explanation can be provided. On the one hand, we note that the mode 3F* is formed of three folds that are grouped whereas they are arranged periodically along the azimuthal direction for the mode 3F. Therefore the bending energy of the mode 3F* is probably only slightly larger than its 3F counterpart. 
On the other hand, when these three folds are grouped, the area swept by the peripheral folds is important, and, as a consequence, yields a substantial work of the pressure forces. 
For large flow velocities, the importance of pressure work becomes relatively more important, and thus the mode 3F* becomes favored compared to the mode 3F.

%
%
In summary, we have studied how a circular flexible sheet deforms when subject to flow-induced loads. A sequence of four deformation modes has been evidenced. For low flow velocities, the sheet deforms cylindrically, while draping patterns appear for larger flow velocities. In this latter case, the sheet deforms into d-cones of different azimuthal symmetries. As the flow velocity is increased, a periodic 2-fold mode is first observed, then a periodic 3-fold mode, and eventually a non-periodic 3-fold mode. The transition between the first three modes has been predicted using a linear model based on energetic arguments and a qualitative explanation has been proposed to explain why the azimuthal symmetry is eventually lost for the largest flow velocities. In addition, we showed that the only parameter that governs the transition between modes is the dimensionless Cauchy number, in agreement with experimental observations.

\begin{acknowledgements}
We  warmly thank A. Boudaoud for insightful discussions and comments. C.~E. acknowledges support from the European Union (fellowship PIOF-GA-2009-252542).
\end{acknowledgements}

\end{document}